\documentstyle[11pt]{article}
\newcommand{\blankline}{\vskip .3cm}
\newcommand{\f}{\begin{equation}}
\newcommand{\ff}{\end{equation}}

\begin{document}
\vfill
\centerline{\LARGE Finite diffeomorphism invariant observables}
\centerline{\LARGE in quantum gravity}

\blankline
\rm
\vskip1cm
\centerline{Lee Smolin${}^*$}
\blankline
 \centerline{\it   Department of Physics, Syracuse University}
 \centerline {\it Syracuse, New York  13244   U.S.A.}
 \vfill
\vfill
\centerline{ABSTRACT}
\blankline
\noindent
Two sets of spatially diffeomorphism invariant operators are
constructed in the loop representation formulation of quantum
gravity.  This is done by coupling general relativity to an anti-
symmetric tensor gauge field and using that field to pick out
sets of surfaces, with boundaries, in the spatial three manifold.
The two sets of observables then measure the areas of these surfaces
and the Wilson loops for the self-dual connection around their
boundaries.  The operators that represent these observables are
finite and background independent when constructed through a
proper regularization procedure.   Furthermore,  the spectra of the
area operators are discrete so that the possible values that one
can obtain by a measurement of the area of a physical surface in
quantum gravity are valued in a discrete set that includes integral
multiples of half the Planck area.

These results make possible the construction of a correspondence
between any three geometry whose curvature is small in Planck
units and a diffeomorphism invariant state of the gravitational
and matter fields.  This correspondence relies on the approximation
of the classical geometry by a piecewise flat Regge manifold, which is
then put in correspondence with a diffeomorphism invariant
state of the gravity-matter system in which the matter fields
specify the faces of the triangulation and the gravitational field
is in an eigenstate of the operators that measure their areas.
\vfill
${}^*$ smolin@suhep.phy.syr.edu
\eject

\section{Introduction}

 The problem of observables is probably the key problem presently
confronting non-perturbative approaches to the quantization of the
gravitational field.  The problem is difficult because it reflects, in two
 different ways, the problem of making sense of a diffeomorphism
invariant quantum field theory.  Already at the classical level, it is
non-trivial to construct and interpret functions of the dynamical
variables
that are invariant under spacetime diffeomorphisms.  While, it is
true, one
can describe in words a certain limited number of them, what is
needed is
much more than this.  First of all, we must have an infinite number
of
observables, corresponding to the number of physical degrees of
freedom of the theory.  Second, to make the transition to the
quantum
theory one must know their Poisson algebra.

When we come to the quantum theory, new problems emerge
because
essentially all diffeomorphism invariant observables involve
products of local field observables, and so are not directly defined
in the representation space of any quantum  field theory.  They must
be defined through a limiting procedure which is analogous to the
regularization procedures of conventional quantum field theories.
Furthermore, it is necessary to confront the fact that none of the
conventional regularization or renormalization procedures can be
applied to this case.  This is because they all depend on the presence
of
a background metric.  It is then necessary to define new
regularization
procedures which may be applied to field theories constructed
without
a background metric.  Additional background structure does come in
the
definitions of the regulated
 operators; what must be shown is that in the
limits
that the regulators are removed the resulting action is finite and
background independent.

In the last year and a half, some
progress\cite{weave,review-lee} has been made on the
problem of observables in the context of a nonperturbative approach
to
quantum gravity based on the Ashtekar variables\cite{Abhay} and
the loop
 representation\cite{carlolee,gambini-loop}\footnote{Reviews of
previous
 work in this direction are found
 in \cite{review-abhay,review-carlo,review-lee}.}.
 This approach is based on taking as a starting point
a quantum kinematical framework which is based on non-Fock
representations\cite{rayner,abhaychris,review-lee} of certain
non-local observable algebras.  It seems
necessary to pick non-Fock representations as the starting point
for the construction of diffeomorphism invariant quantum field
theories to avoid the dependence of the Fock structure on a fixed
background metric.

What has been learned using this approach may be summarized as
follows\cite{weave,review-lee}:

a)   It seems impossible to construct background independent
{\it renormalization} procedures for products of local fields.
This is
because any
local renormalization procedure is ambiguous up to a local density.
As a
consequence, because we are working in a formalism in which the
basic local observable is a frame field, this theory has no operator
which can represent the measurement of the metric at a point.
This further
means that diffeomorphism invariant operators cannot normally be
constructed
from integrals over products of local fields.

b)  Despite this, there are several non-local observables that can be
constructed as operators  acting on kinematical states by a
regularization procedure appropriate to the non-perturbative
theory.
In all of the cases in which a well defined operator exists in the
limit that the regularization is removed, that operator is finite
and background independent.
Among these operators are those that measure the area of any given
surface, the volume of any given region and the spatial integral of
the
norm of any given one form.   By measurements of these
observables the metric can be determined, in spite of the fact that
there is no local operator which can represent the metric.

c)  The connection between background independence and finiteness
is most
likely general, as there are arguments that for any operator that
can be constructed through a   point splitting regularization
procedure of the type
used in \cite{weave,review-lee}, background indepdence
implies finiteness\cite{carloun}.

d)  The spectra of the operators that measure areas and
volumes are discrete series of rational numbers times the
appropriate Planck units.

e)  Using these results, the
semiclassical limit of the theory can be understood.
Given any classical three metric, slowly varying at the
 Planck scale, it is possible to construct a kinematical
quantum state
which has the property that it is an eigenstate of the
above mentioned operators and the eigenvalues agree
with the corresponding classical values in terms of that
metric, up to terms of order of the inverse of the measured
quantity in Planck units.

These results are very encouraging, but they are
subject to an important limitation. They concern the
kinematical level of the theory, which is the original state
space on which the unconstrained quantum theory is
defined.  The physical states, which are those states that
satisfy the Hamiltonian and diffeomorphism constraints,
live in a subspace of this space.

It would then be very desirable to find results analogous to
these holding for physical operators.  In this paper I report
results which bring us significantly closer to that goal.  These
include the construction of a number of operators which
are invariant under spatial diffeomorphisms.
For example, as I will show below,  it is possible to
construct a diffeomorphism invariant operator that
measures the area of surfaces which are picked out by
the values of some dynamical fields.  Just as in the
kinematical case, the spectrum of this operator is discrete
and includes the integral multiples of half the Planck area.

The basic idea on which these results are based is to use
matter fields to define physical reference frames and then
use these to construct diffeomorphism invariant observables.
Of course, the idea of using matter fields
to specify dynamically a coordinate system  is very
old.  It goes back to
Einstein\cite{albert-snail}, who pointed out that in order to realise
the operational definition of lengths and times in terms
of rulers and clocks in general relativity, it was necessary
to consider the whole dynamical system of the rulers
and clocks, together with the gravitational field.  The
application of this idea to the quantum theory was first
discussed by DeWitt\cite{bryce-snail}, and has been recently revived
by Rovelli\cite{carlo-matter,carloobserves} and by Kuchar and
Torre\cite{karel-matter}.

In a paper closely related to this one, Rovelli has used a scalar field
to pick out a set of surfaces whose areas are then measured.
In this paper I take for the matter field an
antisymmetric tensor gauge field, with dynamics as first written
down by Kalb and Ramond\cite{kalbramond}.   There are two
reasons
for this.  First, as we will see below,
the coupling of the antisymmetric
tensor gauge field to gravity is particularly simple in the
Ashtekar formalism, which allows us to hope that it
will be possible to get results about physical observables,
which must commute also with the Hamiltonian constraint.
The second
reason is that, as I will describe, the configurations of
the Kalb-Ramond field can be associated
with open surfaces, which has certain advantages.

Now, it is clear on the kinematical level that if one measures the
area of every two dimensional surface one determines the
spatial metric
completely.  One can then imagine that if one has a finite, but
arbitrarily large, set of surfaces, one can use measurements of
their areas to make a partial measurement of the metric.  Such
an arrangement can serve as a model of an apparatus that might
be used to measure the gravitational field, because indeed, any
real physical measuring device returns a finite amount of
information and thus makes only a partial measurement of a
quantum field.  As I discuss below, there are a number of results
and lessons that can be learned about measurement theory for
quantum gravity by using a finite collection of surfaces as a model
of a measuring apparatus.

Once we have a set of finite spatially diffeomorphism invariant
operators,
defined by using matter fields as a quantum reference frame,
it is interesting to try to employ the same strategy to construct
physical observables\footnote{Note that, as pointed out by
Rovelli\cite{carloobserves}, there is a model diffeomorphism
invariant
theory in $3+1$ dimensions whose Hamiltonian constraints
is proportional to a linear combination of the gauge and
diffeomorphism constraints.  This is the Husain-Kuchar
model\cite{husainkuchar}, which corresponds, at the classical level,
to a
limit of general relativity in which the speed of light has
been taken to zero.   The physical state space of this model
is just the space of diffeomorphism invariant states of quantum
gravity and the
physical inner product is known and takes a simple form in
the loop representation\cite{husainkuchar}.  The observables I
construct
here are then  examples of  physical observable if we
adjoin to the Husain-Kuchar model the antisymmetric
gauge field.}.  One can add to the theory
additional matter degrees
of freedom which can represent physical clocks and use these to
construct
operators which commute with the Hamiltonian constraint but
describe
measurements made at particular times as measured by the physical
clock.  This idea is developed in \cite{me-dieter}.  Furthermore, once one
has
physical operators that correspond to measurements localized in
space
and time by the use of matter fields to form a spacetime reference
system, it is possible to give a formulation of a measurement theory
which may be applied to quantum cosmology.  A sketch of such a
measurement theory is also developed there.

This paper is organized as follows.  In the next section I show how to
couple an antisymmetric tensor gauge field to gravity.  This is
followed
by section 3 in which I show how to quantize the tensor gauge field
in terms of a surface representation that is closely analogous to
the abelian loop representation for Maxwell
theories\footnote{Rodolfo
Gambini has kindly
informed me that many of the results of this section
were found previously
in \cite{rodolfo-antisym}.}.  I then show how to combine these
results
with the loop representaion of quantum gravity and how to construct
diffeomorphism
invariant states of the coupled system.  Here it is also shown how
to construct
the diffeomorphism invariant operator that measures the area of the
surface picked out by the quantum state of the antisymmetric tensor
gauge field.   In section 4 I consider a
straightforward extension of these results in which certain degrees of
freedom are added which result in quantum states which are labled
by
open rather than closed surfaces. In the next section, which is section
5, I show how the
matter field may also be used to construct a diffeomorphism
invariant
loop operator.  This has the effect of adding
Wilson loops of the left-handed
spacetime connection around the edges of the surfaces picked out by
the
quantum states of the matter.  These results are then used
in section 6, where I show
how to construct  quantum reference systems by combining the
surfaces
from a number of independent matter degrees of freedom to
construct
simplicial complexes.  We find here a very interesting
correspondence
between certain peicewise-flat Regge manifolds and the elements of
a
basis of diffeomorphism invariant states of the coupled matter-
gravity
system. I also sketch an approach to a measurement theory
for quantum gravity which is based on the
results described here\cite{me-dieter}.
 Finally, the implications of these results
are the subject of the conclusion.

\section{Coupling an antisymmetric tensor gauge theory to gravity}

An antisymmetric tensor gauge field\cite{kalbramond} is a two form,
$C_{ab}=-C_{ba}$ subject to a gauge transformation
generated by a one form $\Lambda_a$ by,
\f
\delta C_{ab} = d\Lambda _{ab} .
\ff
It's field strength is a three form which will be
denoted $W_{abc}=dC_{abc}$.    The contribution
to the action for these fields coupled
to gravity is, in analogy with electromagnetism,
\f
S_C= {k \over 4}
 \int d^4x \sqrt{g}g^{ad}g^{be}g^{cf}W_{abc}W_{def}
\ff
where $k$ is a coupling constant with dimensions of inverse action
and all the quantities are,
just for moment, four dimensional.

In the Hamiltonian
theory\footnote{From now on we consider that the indices
$a,b,c,...$ are spatial indices, while indices $i,j,k,...$ will be
internal $SO(3)$ indices.  Densities, as usual, are sometimes, but
not always, indicated by a tilde.}   its conjugate momenta is
given by $\tilde{\pi}^{ab}=-\tilde{\pi}^{ba}$ so that
\f
\{ C_{ab} (x) , \tilde{\pi}^{cd} (y) \} =
\delta^{[c}_a \delta^{d]}_b \delta^3 (y,x)
\ff
where the delta function is understood to be a density with
the weight on the first entry and $\tilde{\pi}^{cd}$ is also a density.
We will also find it convenient to work with the dualized fields
\f
\tilde{W}^*={1 \over 3!}\epsilon^{abc}W_{abc}
\ff
and
\f
\pi^*_a={1 \over 2} \epsilon_{abc}\tilde{\pi}^{bc}.
\ff
The gauge transform (1) is then generated by the constraint
\f
G=\partial_c \tilde{\pi}^{cd} =d\pi^*_{cd}=0
\ff

This field can be coupled to gravity in the Ashtekar formalism by
adding to
the Hamiltonian constraint the term
\f
{\cal C}_{matter}= { k \over 2}  (\tilde{W}^*)^2  + {1 \over 2k }
\pi^*_a \pi^*_b \tilde{\tilde{q}}^{ab}
\ff
and adding to the diffeomorphism constraint the term
\f
{\cal D}^{matter}_a = \pi^*_a \tilde{W}^*.
\ff

We may note that the term added to the Hamiltonian
constraint is naturally a density of weight two, so that
it is polynomial without the necessity of changing the
weight of the constraint by multiplying by a power of
the determinant of the metric, as is necessary in
Maxwell or Yang-Mills theory\cite{ART}.

The antisymmetric tensor gauge field can be understood
to be a theory of surfaces in three dimensions in the same
sense that Maxwell theory is a theory of the Faraday flux
lines\cite{carloted}.  By (6) there is a scalar field $\phi $
such that locally\footnote{Global considerations play no role
in this paper.}
\f
\pi^*_a =d\phi_a
\ff
The equipotential surfaces of $\phi $ define a set of surfaces
which are the analogues of the Faraday lines of
electromagnetism.  Further, any two dimensional
surface $\cal S$ defines a  distributional configuration of the
$\pi^{ab}$ by,
\f
\pi^{ab}_{\cal S} (x) = \int d^2S^{ab}(\sigma ) \delta^3 (x ,{\cal S}
(\sigma ))
\ff
here $\sigma$ are  coordinates on the surface.
Note that $\pi^{ab}_{\cal S}$ is automatically divergence free.
These are completely analogous to the distributional
configurations of the electric field\cite{review-lee,lee-cs}
that may be associated to  a curve $\gamma$ by
\f
\tilde{E}^a_{\gamma} (x) = \int d\gamma^a (s) \delta^3 (x,\gamma (s))
\ff
We can define a diffeomorphism invariant observable which depends
on
both the metric and $\tilde{\pi}^{cd}$ which has the property that
when
$\tilde{\pi}^{cd}$
has such a distributional configuration it measures the area of that
surface.
This can be defined either by
generalizing the definition of the area
observable\cite{weave,review-lee} or more directly by
\f
A(\pi, \tilde{E}) \equiv  Q (\pi, \tilde{E}) = \int
\sqrt{\tilde{\tilde{q}}^{ab} \pi^*_a \pi^*_b}
\ff
An equivalent expression for this is given by
\f
A(\pi ,\tilde{E} ) = \lim_{N \rightarrow \infty} \sum_{N=1}^N
\sqrt{A^2_{approx} [{\cal R}_i]}
\ff
where space has been partitioned into $N$ regions ${\cal R}_i$ such
that in the limit $N \rightarrow \infty$ the regions all shrink to
points.  Here, the
observable that is measured on each region is defined
by\footnote{Here $T^{ab}(x,y)$ is defined as follows\cite{review-lee}.
Let there
be a procedure, based on a background flat metric, to associate a
circle
$\gamma_{x,y}$ to every two points in the three manifold $\Sigma$.
Then define
$T^{ab}(x,y) \equiv TrU_{\gamma_{xy}}(y,x) \tilde{E}^a (x)
U_{\gamma_{xy}}(x,y) \tilde{E}^b (y)$, where
$U_\gamma (x,y) \equiv Pexp G \int_\gamma A $ is parallel
transport along the curve $\gamma$ from $x$ to $y$.},
\f
A^2_{approx} [{\cal R}] \equiv \int_{\cal R} d^3 x \int_{\cal R} d^3y
\ \ T^{ab} (x,y) \pi^*_a (x) \pi^*_b (y)
\ff

To show the equivalence between these two expressions, we may
start with
(12) and regulate it the way it is done in the quantum theory by
introducing a background euclidean coordinate system and a set of
test fields $f_{\epsilon} (x,y)$ by
\f
f_{\epsilon}(x,y) \equiv {\sqrt{q(x)} \over \epsilon^3 }
\theta [{\epsilon \over 2}-|x^1-y^1 |]
\theta [{\epsilon \over 2} -|x^2-y^2 |]\theta [{\epsilon \over 2}-|x^3-
y^3 |]
\ff
In these coordinates
\f
\lim_{\epsilon \rightarrow 0} f_{\epsilon} (x,y) = \delta^3 (x,y)
\ff
We can then write
\f
A(\pi , \tilde{E}) = Q(\pi,\tilde{E}) = \lim_{\epsilon \rightarrow 0}
\int d^3x
\sqrt{\int d^3 y \int d^3z T^{ab}(y,z) \pi^*_a (y) \pi^*_b (z)
f_\epsilon(x,y) f_\epsilon (x,z)}
\ff
When the expression inside the square root is slowly
varying in $x$ we can
reexpress it in the following way.  We divide space into
regions ${\cal R}_i$ which are cubes of volume
$\epsilon^3$ centered on the points
$x_i = (n\epsilon , m\epsilon , p \epsilon )$ for $n,m,p$ integers.  We
then
write,
\begin{eqnarray}
A(\pi , \tilde{E}) &= &\lim_{\epsilon \rightarrow 0} \sum_i
\epsilon^3
\sqrt{\int d^3 y \int d^3z T^{ab}(y,z) \pi^*_a (y) \pi^*_b (z)
f_\epsilon(y,x_i) f_\epsilon (z,x_i)}   \nonumber  \\
&=&\lim_{N \rightarrow \infty} \sum_{N=1}^N \sqrt{A^2_{approx}
[{\cal R}_i]}
\end{eqnarray}

If we now plug into these expressions the distributional
form (10) it is straightforward to show that
\f
A(\pi_{\cal S} , \tilde{E} )= \int_{\cal S} \sqrt{h}
\ff
where $h$ is the determinant of the metric of the two
surface, which is given
by $h=\tilde{\tilde{q}}^{ab}n_a n_b$ where $n^a$ is the
unit normal of the surface.

\section{Quantization}

It is straightforward to construct an algebra of loops and
closed surfaces to
coordinatize the gauge constraint surface, corresponding to
the imposition of both $G=0$ and the $SU(2)$ Gauss's
law of the gravitaional fields.   We may associate
to every closed surface $\cal S$ a gauge invariant observable,
\f
T[{\cal S}] \equiv e^{\imath k \int_{\cal S} C} \ \ \ .
\ff
Conjugate to $T[{\cal S}]$ we have the observables
$\tilde{\pi}^{ab} (x)$
which
 satisfy the algebra,
\f
\{ T[{\cal S}] , \tilde{\pi}^{ab} (x) \} =
\imath k \int d^2 {\cal S}^{ab}
(\sigma )
\delta^3 (x , {\cal S}(\sigma )) T[{\cal S}]
\ff

We would like now to construct a representation of this algebra
as an algebra of operators.  We can construct a surface
representation
in which the states are functions of a set of closed surfaces
$\Psi [\{ {\cal S} \} ]$ \footnote{Such a representation was first
constructed in \cite{rodolfo-antisym}.}.  In order
to implement the abelian gauge invariance we require
that these states are invariant under reparametrization
invariance and satisfy two relations.  First, we require that
\f
\Psi [ {\cal S} \circ {\cal S}^\prime ] = \Psi [ {\cal S} \cup {\cal
S}^\prime ]
\ff
where ${\cal S} $ and ${\cal S}^\prime$ are any two
surfaces that touch at one point and
${\cal S} \circ {\cal S}^\prime$ is the surface made
by combining them.  Second, we require that
$\Psi [ {\cal S}] = \Psi [ {\cal S}^\prime] $ whenever
$e^{\int_{\cal S} F }= e^{\int_{{\cal S}^\prime} F }$ for every
two form $F$.
We then define the representation by,
\f
\hat{T}[{\cal S}^\prime ] \Psi [ {\cal S}]
= \Psi [ {\cal S}^\prime \cup {\cal S}]
\ff
and
\f
\hat{\pi}^{ab} (x) \Psi [{\cal S}] = \hbar k \int d^2 {\cal S}^{ab}
(\sigma )
\delta^3 (x, {\cal S}(\sigma ) ) \Psi [{\cal S}]  .
\ff
 It then follows that the operators satisfy
\f
[ \hat{T}[{\cal S}] , \hat{\pi}^{ab} (x) ] = -\hbar k  \int d^2
{\cal S}^{ab} (\sigma )
\delta^3 (x, {\cal S}(\sigma ) )
\hat{T}[{\cal S}]
\ff

We should now say a word about dimensions.    In order
that the interpretation of $A[\pi ,\tilde{E}]$ as an area
work, it is necessary that
$\tilde{\pi}^{bc}$ have dimensions of inverse length,
from which it follows from (7) that $k$ have dimensions
inverse to $\hbar$ and that the dimensions of
$C_{ab}$ are $mass/length$.
This choice is consistent with both the Poisson bracket and
the requirement that the exponenent in (20) be dimensionless.

We now bring gravity in via the standard loop
representation\cite{carlolee}.  The states are then
functions, $\Psi [\alpha , {\cal S}]$, of loops and surfaces.
We may introduce a set of bra's $<\alpha , {\cal S}|$ labled
by loops and surfaces so that,
\f
\Psi [\alpha , {\cal S}] =<\alpha , {\cal S}|\Psi >
\ff
We then want to express the area observable (13) as a
diffeomorphism invariant operator and show that it
does indeed measure areas.  It is straightforward to
show that the bras \newline
$<\alpha , {\cal S}|$ are, for
nonintersecting loops $\alpha$, eigenstates of
the operator $\hat{A}$.
This operator may be constructed by using the expression (13) as a
regularization, in the way described in detail in \cite{review-lee}.
A straightforward calculation shows that for the case
of nonintersecting loops\footnote{The appearance of the
Planck area is due to the presence of $G$ in the definition of the
parallel propogators for the Ashtekar connection, $A$, as in the
kinematical case \cite{weave,review-lee}.  This is due to the fact
that it is $GA_a$ that has dimensions of inverse length.  The
dimensionality of the gravitational constant
thus manifests itself in the
appearance of the Planck area in the operator algebra for quantum
gravity.}
\f
<\alpha , {\cal S}| \hat{A}^2_{approx} [{\cal R}] =
({ \hbar kl_{Planck}^2 \over 2 })^2
I[\alpha , {\cal S} \cap {\cal R} ]^2 <\alpha , {\cal S}|
\ff
where $I[\gamma , {\cal S}] $ is the  intersection number given by,
\f
I[\gamma , {\cal S}] \equiv \int d\gamma ^a (s) \int d^2 {\cal S}^{bc}
(\sigma ) \delta^3 ({\cal S}(\sigma ) , \gamma (s)) \epsilon_{abc}
\ff
and
where ${\cal S} \cap {\cal R} $ means the part of the surface that lies
inside the region.  It then follows from (13) that
\f
<\alpha , {\cal S}|  \hat{A} = { \hbar k l_{Planck}^2 \over 2 }
I^+[\alpha , {\cal S}]  <\alpha , {\cal S}|
\ff
where $I^+ [\alpha ,{\cal S}]$ represents the positive definite
unoriented
intersection number which simply counts all intersections
positively.
Thus, we see that the operator assigns to the surface
an area which is given by
$\hbar k l_{Planck}^2 / 2 $ times the number of
intersections of the loop with
the surface\footnote{When the loop $\alpha$ has intersections
at the surface ${\cal S}$ there are additional terms in the
action of the area operator\cite{review-lee}.}.

The action (29) of the area
operator is diffeomorphism invariant, because the surface is
picked
out by the configuration of the field. (One may check that
this is also the case when the loop has an intersection at
the surface.)  The operator
is then well defined acting on
states of the form
\f
\Phi [\{ \alpha , {\cal S} \} ] = < \{ \alpha , {\cal S} \} |\Phi >
\ff
where $\{ ...\}$ denotes equivalence classes
under diffeomorphisms.
On the space of diffeomorphism invariant
states we can impose the natural inner product.  Again, restricted
to the case of nonintersecting loops this must have the form,
\f
<\{ \alpha , { \cal S} \} | \{ \beta , {\cal S}^{\prime} \} >
= \delta_{\{ \alpha , {\cal S} \} \{ \beta , {\cal S}^\prime \} }
\ff
where the delta function is a kronocker delta of knot classes.
The definition of the inner product on intersecting loops may
be obtained by imposing reality conditions.  The complete set
of reality conditions at the diffeomorphism invariant level is
not known, but it is known that an inner product that
satisfies (31) is consistent with the requirment that
$\hat A$ be a hermitian
operator.

We may then conclude that the  spectrum
of $\hat A$ is discrete.
It consists first of the series integer multiples of
$\hbar k l_{Planck}^2 / 2$,
together with a discrete series of other
eigenvalues that come from eigenstates
similar to those discussed in \cite{review-lee} in which the
loops have intersections at
the surfaces.

FInally, may then note that if we require that the
diffeomorphism invariant operator
yield, when acting on kinematical states of the
kind described in \cite{weave,review-lee}, the
same areas as the kinematical area operator,
we get the condition that,
\f
k={1 \over \hbar}
\ff
With its coupling thus set by $\hbar$, the
antisymmetric tensor gauge field is
then in a sense purely a quantum phenomena.

\section{Adding a boundary}

In the next section I am going to make use of the
quantum antisymmetric
tensor gauge field to construct a quantum reference
system for measuring
the diffeomorphism invariant states of the gravitational field.
For this and
other purposes, it is convenient to have states which are
labled by open surfaces in addition to those described in the
previous
section in
which gauge invariance restricts
the surfaces to be closed.  As I will now describe, there is a
very simple way
to do this, which is analogous to the Abelian Higgs model and was
described first by Kalb and Ramond\cite{kalbramond}.
We will see that
by coupling the $C_{ab}$ field to a vector field in a way that
preserves the
gauge invariance (1) we open up the possibility for our
surfaces to have
boundaries.

Let us consider then adding to the system
described by (2) an ordinary
abelian gauge  field,
$b_a$, with an Abelian gauge group given by
\f
\delta b_a = \partial_a \phi
\ff
where $\phi$ is a scalar field.  We may couple this
field to the Abelian
tensor gauge field by
supplementing the gauge transformations (1) by
\f
\delta b_a = \Lambda_a
\ff
Thus, we see that this vector field can be set to zero
by a gauge transform.
A field strength for $b_a$ that is invariant under
both abelian gauge
invariances may be defined by
\f
F_{ab} = db_{ab} -C_{ab}
\ff

To define the dynamics of this coupled system we add
to the action the term
\f
S_{b}={k\over 4}  \int d^4x \sqrt{g}g^{ab} g^{cd} F_{ac} F_{bd}
\ff
We can define a constained Hamiltonian system by
adding (2) and (36) to the
gravitational action.  If the conjugate momenta to the
$b_a$ are labled as
$ \tilde{p}^a $ the  diffeomorphism and gauge
constraints (8) and (6) are now
\f
D_a = \tilde{W}^* \pi_a^* + \tilde{p}^c F_{ac}
\ff
and
\f
G^a = \partial_b \tilde{\pi}^{ab} + \tilde{p}^a
\ff
The Hamiltonian constraint has additional terms,
which are given by
\f
{1 \over 2k} \tilde{p}^a\tilde{p}^bq_{ab} +
{k \over 2} det(q) F_{ac}F_{bd}q^{ab}q^{cd}
\ff
Note that the new terms are non-polynomial, when expressed
in terms of the canonical variables $\tilde{E}^a_i$, as in the
case of the Maxwell
and Yang-Mills theories\cite{ART}. (As in that case this can be
remedied by
multiplying through by
$det(q_{ab})$.)  Finally, there is a new constraint,
\f
g=\partial_c \tilde{p}^c
\ff
which generates (33).  This, however, is not independent of (38) as
\f
\partial_a G^a = g
\ff
As a result, there are now three independent gauge
constraints and six
each of canonical coordinates and momenta.  Thus, the
theory now has three
degrees of freedom per point.  The two additional
degrees of freedom are
reflected in the fact that in addition to the one gauge
invariant field $\tilde{W}^*$,
we now have the gauge invariant two form $F_{ab}$.
Three of these four
gauge invariant degrees of freedom are independent,
because we have
\f
dF_{abc} = -W_{abc}
\ff
As a result, we can associate gauge invariant observables
to each open surface $\cal S$.  This is given by
\f
T[ {\cal S} ] = e^{{1 \over k}  \int_{\cal S} F}
\ff
The poisson brackets of this with the canonical
momenta $\tilde{\pi}^{ab}$ and
$p^a$ are given by
\f
\{ \tilde{\pi}^{ab} (x) , T[{\cal S}] \} = -{\imath \over k} \int d^2S^{ab}
(\sigma )
\delta^x (x , S(\sigma ) ) T[{\cal S}]
\ff
\f
\{ p^a (x) , T[{\cal S}] \} = {\imath \over k} \int ds
\delta^x (x , \partial S(s ) ) \dot{\partial {\cal S}}(s)    T[{\cal S}]
\ff

The surface representation defined by (23) and (24) can be
extended in the obvious
way.  The arguements of the states are now open
surfaces and the
obvious combination laws of surfaces hold.  In addition to
(24), which still
holds, there is the operator
\f
\hat p^a (x) \Psi [{\cal S}] = {\hbar \over k}
 \int ds
\delta^x (x , \partial S(s ) ) \dot{\partial {\cal S}}(s)   \Psi [{\cal S}] .
\ff

Finally, one may check that the gravitational degrees of freedom
may be added and the area operator defined, so that all the
results of the previous section extend naturally to the surface
representation with boundaries.

\section{A diffeomorphism invariant loop operator}

Given that the matter fields specify a set of surfaces with boundaries,
we may imagine constructing a diffeomorphism invariant
holonomy operator, analogous to the $T^0[\alpha ]$ operators of the
kinematical theory, in which the loop $\alpha$ is given by the
boundary $\partial {\cal S}_I$ of the surface determined by the
$I$'th
matter field.

To do this we first need to construct an appropriate diffeomorphism
invariant classical observable that will measure
the holonomy of $A_a^i$ on such loops.
This can be done by using the fact that
$\tilde{p}^a$, by virtue of its being
a divergence free vector density, defines a
congruence of flows.  These flows may be labeled by a two
dimensional
coordinate $\sigma^\alpha$, with $\alpha =1,2$, which may be
considered to
be scalar fields on $\Sigma$ that are constant along the flows.
The idea is to
define a generalization of the trace of the
holonomy from a curve to a congruence
by taking the infinite product of the traces of the holonomies over
each
curve in the congruence.  This may be done in the following way.

Each divergence free vector density may be written as a two form in
terms of the two functions $\sigma^\alpha $ as \cite{carloted},
\f
p^*_{ab }= (d \sigma^1 \wedge d \sigma^2 )_{ab}
\ff
where the $\sigma^\alpha$ are two scalar functions that are constant
along
the curves of the congruences and so may be
taken to label them.  The curves of the congruences may be written
as
$\gamma^a_p (\sigma , s) $ and satisfy,
\f
\dot{\gamma}^a_p(\sigma , s) \equiv {d \gamma^a_p (\sigma , s)
\over ds}
= \tilde{p}^a
\ff
We may note that because each $\tilde{p}^a$ is
divergence free it is the case that through every point $x$ of
$\Sigma $ there passes at most one curve of
the congruence.  We will denote this curve by $\gamma_{p}(x)$.
We may
take as a convention that if no curve of the congruence passses
through $x$
we have $\gamma_{p}(x)=x$, which is just the degenerate
curve whose image is just the point $x$.  Further, note that we
assume
that either appropriate boundary conditions have been imposed
which
fix the gauge at the boundary or we are working in the context of a
closed manifold $\Sigma$, for which the curves $\gamma_p (x)$ are
closed.

We may then define a
classical observable   which is the trace of
the holonomy of the connection around the curve $\gamma_{p}(x)$.
\f
W[p,A](x)=  Tr U_{\gamma_{p}(x)}
\ff
where $U_\gamma$ is the usual path ordered holonomy of $A$ on
the curve
$\gamma$.  We may note that the observable $W[p,A](x)$
transforms as a scalar field.

We may now write a diffeomorphism and gauge invariant observable
which
is
\f
T[p,A] \equiv e^{\int d\sigma_1 d\sigma_2 LN
TrU_{\gamma_{p,\sigma}}}
\ff
To show that this is indeed diffeomorphism invariant, as well
as to facilitate
expressing it as a quantum operator, it is useful to rewrite
it in the following
way.  Let ${\cal S}_{\tilde{p}}$ be an arbitrary two surface
subject only to the condition that it intersects each curve
in the congruence determined by
$\tilde{p}^a$ exactly once
so that  ${\cal I}[\gamma_{p,\sigma},{\cal S}_{\tilde{p}}] =1 $.
Then we may write
\f
T[p,A]= e^{\int d^2 {\cal S}^{ab}_{\tilde{p}}
p^*_{ab} LN W[p,A]      }
\ff
The diffeomorphism invariance of this observable is
now manifest.
To see why this form may be translated to a
diffeomorphism
invariant quantum
operator, we may note that it reduces
to a simple form if we plug in for $\tilde{p}^a$
the distributional
divergence free vector density
\f
\tilde{p}^a_\alpha \equiv \int ds \delta^3 (x, \alpha (s))
\dot{\alpha}^a (s) .
\ff
It is then not hard to show that
\f
T[p_{\alpha} ,A] = Tr P e^{\int_\alpha A } = T[\alpha ]
\ff

We may now define a quantum operator $\hat{T}$
corresponding to (51) by replacing
$\tilde{p}^a$ with the corresponding operator (46),
\f
\hat{T} = T[\hat{p},\hat{A}] .
\ff
As all the operators in its definition commute, there is
no ordering issue.  It is then straightforward to show that
\f
< \{ S , \gamma \} | \hat{T}=
< \{ S , \gamma \cup \partial S\} |
\ff

That is, the action of the $\hat{p}$ operators in
(51) is by (46) to
turn the operator into a loop operator for the holonomy
around the
surface.  The result is that what the operator $\hat{T}$
does is to
add a loop
to the diffeomorphism equivalence
class $\{ S, \gamma \}$ which
is exactly the boundary of the  surface.
Thus, we have succeeding
in constructing a diffeomorphism invariant loop operator.

I close this section by noting two extensions of this result.  First,
if one considers the case of Maxwell-Einstein theory, where both
fields are treated in the loop
representation\footnote{The loop representation for Maxwell
fields is described in\cite{abhaycarlo,gambini} and the coupling
of Maxwell to gravity in the Ashtekar formalism is described in
\cite{ART}.}, one has an analogous
operator, where $\tilde{p}^a$ should be taken to be just the electric
field.  In this case, if a diffeomorphism invariant quantum state is
given by $\Psi [\{ \alpha , \gamma \}]$ where $\alpha $ are the
abelian
loops that represent the electromagnetic field and $\gamma$ are
the loops that represent the gravitational field and
$\hat{T}$
is
the operator just described we have
\f
\hat{T} \Psi [\{ \alpha , \gamma \} ]=
\Psi [\{ \alpha , \gamma \cup \alpha \} ] .
\ff
That is, the operator puts a loop of the self-dual graviational
connection over each loop of the electromagnetic potential.

Second, all of the considerations of this paper apply to the system
which is gotten by taking the $G \rightarrow 0$ limit of general
relativity in the Ashtekar formalism\cite{Gtozero}.  This limit yields
a chirally asymmetric theory whose phase space consists
of all self-dual configurations together with their linearized
anti-self-dual perturbations.  In this case there are operators
$\tilde{e}^a_i$ and $A_a^i$ which are cannonically conjugate, but the
internal gauge symmetry is the abelian $U(1)^3$ reduction of the
internal
$SU(2)$ gauge symmetry.  One then has an operator analogous
to (51), which is just
\f
T[e,A]_{G \rightarrow 0 } \equiv
e^{\int_\sigma \tilde{e}^a_i A_a^i}
\ff
The corresponding quantum operator
has the effect of increasing the winding numbers of loops that
are already present. It is also interesting to note that in this
case, $T[e,A]_{G \rightarrow 0 } $ commutes with the Hamiltonian
constraint, so that it is actually a constant of the motion
\cite{Gtozero}.

\section{A quantum reference system}

I would now like to describe how the preceding
results can be used to construct a physical interpretation
of a very large class of diffeomorphism invariant states.
As I mentioned in the introduction,
the idea of using matter fields to provide a dynamically
defined coordinate system with respect to which a
diffeomorphism invariant interpretation of the
gravitatational fields can be defined
 was introduced into quantum gravity by De Witt's paper
\cite{bryce-snail} in which he applied the Bohr-Rosenfeld analysis
to the problem of the measurability of the quantum gravitational
field.
It is interesting to note that in this paper DeWitt
concluded that it was impossible to make measurements in
quantum gravity that resolved distances shorter than the
Planck scale.  The results of the present paper reinforce
this result and add to it two  important dimensions: first
that, at least in one approach, it is impossible to measure
things smaller than Planck scales because the fundamental
geometrical quantities are quantized in Planck units and
second, that it is areas and volumes\footnote{For the volume
operator see \cite{review-lee}.}, and not lengths,
whose measurements are so quantized.

Let us consider, for simplicity, that there are many species
of antisymmetric-tensor gauge fields, $(C_{ab}^I, b_a^I)$,
labled by the index $I=1,...,N$, where $N$ can be taken
arbitrarily large.  This is a harmless assumption as long
as we are concerned only with spatially diffeomorphism
invariant states.  I will come back to this point in the conclusion.

By the straightforward extension of all the results of
the previous section, quantum states are now functions
of $N$ surfaces, ${\cal S}_I$, so that
\f
\Psi [\{ \gamma , {\cal S}_I \} ] = < \{ \gamma , {\cal S}_I \} | \Psi >
\ff

We may note that the space of diffeomorphism equivalence classes,
$\{ \gamma , {\cal S}_I \}$ of loops and $N$  labeled open
surfaces is countable\footnote{Note that each surface may be
disconnected.}.  The diffeomorphism invariant
state space of quantum gravity coupled to the $N$
antisymmetric tensor gauge fields then has a
countable basis given
by
\f
\Psi_{ \{ \alpha  , {\cal S}^\prime_I \}  }
[\{ \gamma , {\cal S}_I \} ] =
\delta_{ \{ \alpha  , {\cal S}^\prime_I \} \{ \gamma , {\cal S}_I \}  }
\ff
in the case that the loop $\gamma$ is not
self-intersecting.  In the intersecting case, the
form of the basis elements is more complicated because of
the presence of the non-trivial relations among intersecting loops
which result from the identities satisfied by $SU(2)$ holonomies.
For the kinematical case, these relations, and the effect on
the characteristic inner product are described in \cite{review-lee}.
For the present, diffeomorphism invariant, case they have not
yet been completely worked out.   However, for the results I
will describe below it is sufficient to restrict attention to
diffeomorphism equivalence classes involving
only non-intersecting loops.

Let us now consider a particular subspace of states of this
form which are defined in the following way.  Let us
consider a particular triangulation of the three
manifold, $\Sigma$, labled $\cal T$.  It consists of  some
number,
$M$, of  tetrahedra, labled ${\cal T}_\alpha$, where
$\alpha =1,...,M$, that have been
joined by identifying faces.   Let us call the faces
${\cal F}_I$ and let us consider only $\cal T$ that contain
exactly $N$ faces so that $I=1,...,N$.
The idea is then to use this triangulation
 to construct a quantum coordinate system by identifying
each face
${\cal F}_I$ with the surface ${\cal S}_I$ which is an
excitation of the $I$'th
matter field.

We do this in the following way.
For each such triangulation of $\Sigma$ we can consider a
subspace of states, which I will denote
${\cal S}_{\cal T}$, which consists of all
states that have the form
\f
\Psi [\{ \gamma , {\cal S}_I \} ] =
\delta_{ \{  {\cal F}_I \} \{ {\cal S}_I \} }
\psi [\{ \gamma , {\cal S}_I \} ] .
\ff
where the $\delta_{ \{  {\cal F}_I \} \{ {\cal S}_I \} } $
is, again, a topological Kronocker delta that is
equal to one if and only if
each surface ${\cal S}_I$ can be
put in correspondence with the face
 ${\cal F}_I $ such that all the topological
relations among the surfaces are preserved.
Such
an  arrangement of surfaces can be taken to
constitute a quantum reference frame.  The
states in ${\cal S}_{\cal T}$ can
then take any value as we vary over the
countable  set of diffeomorphism equivalence classes
in which the loops are knotted and linked with the
surfaces in $\cal T$ and with each other in all
possible diffeomorphically inequivalent ways.

If we impose an additional restriction, we can make a
correspondence between a basis for
${\cal S}_{\cal T}$ and a countable set of peicewise
flat three dimensional manifolds based on the
simplicial complex $\cal T$.   This restriction is
the following: in any three dimensional simplical
complex the number of faces, $F({\cal T})$ is greater
than or equal to the number of links,
$L({\cal T})$ \cite{alan-personal}.
For a reason that will be clear in
a moment, let us
restrict attention to $\cal T$ such that
$F({\cal T})=L({\cal T})$.

Let us then consider the characteristic basis for
${\cal S}_{\cal T}$ given by
(59) with $\{ {\cal S}_I \} = {\cal T}$.  In any such
state we may then associate a definite value for
the area of each face
in ${\cal T}$, which is given by the eigenvalue
of ${\hat A}^I$.

We may then associate to each set of areas
${\cal A}^I$ a piecewise flat manifold, which I will
call ${\cal M}_{  \{ {\cal A}^I , {\cal S}_I \} }$, which is
composed of flat tetrahedra glued together with the
topology of $\cal T$ such that the areas of the faces are
given by the ${\cal A}^I$.  We know that generically
this can be done, because such piecewise geometries are
determined by the edge lengths of the triangulation,
and we have assumed that the number of edges in
$\cal T$ is equal to the number of faces.  Thus, we
may in general invert the $N$ relations between the
edge lengths and the areas of the faces to find the
edge lengths.  However, when doing this, we need to
be careful of one point, which the following.

Note that
we have chosen
the signs while taking the square root in (13) so that all
areas are positive.  However, if we consider a tetrahedron
in $\cal T$, there is no reason for the areas of the four sides
to satisfy the tetrahedral identities, which imply that the sum
of the areas of any three sides is greater than the area of the
fourth side.  This means that we cannot associate to  each
tetradhedra of $\cal T$ a metrically flat tetrahedra, if
we require
that the signature of its metric be positive definite.
Instead,
we must associate a flat metric of either positive or
negative signature, depending
on whether or not the classical tetrahedral identities are
satisfied.   Thus, whether a particular surface of a
particular tetrahedra is spacelike,
timelike or null depends on how the identities are satisfied
in that tetrahedra.

However, each surface bounds two tetrahedra and there is no
reason that the signiture of the metric may not change as the
surface is crossed.  Thus, a surface may be, for example,
 timelike with respect
to its imbedding in one of the tetrahedra it bounds, and
spacelike in another, as long as the absolute values of the
areas are the same.  Similarly, when the edge lengths are
determined from the areas it is necessary to use the
appropriate formula for each tetrahedra, which depends on
the signature of the metric in that tetrahedra.

Thus, the result is that the piecewise flat manifold
${\cal M}_{  \{ {\cal A}^I , {\cal S}_I \} }$ that is determined
from the $N$ areas ${\cal A}^I$ in general contains
flat tetrahedra with different signatures, patched together
so that the absolute values of the areas match.  Additional
conditions, which are precisely the tetrahedral identities,
must be satisfied if the geometry of
${\cal M}_{  \{ {\cal A}^I , {\cal S}_I \} }$ is to correspond
to a positive definite metric on $\Sigma$.

We may note also that the
 correspondence between the piecewise flat
three geometry,
${\cal M}_{  \{ {\cal A}^I , {\cal S}_I \} }$,
and the diffeomorphism equivalence classes
$\{ \gamma , {\cal S}_I \} $ is
not one to one.  Given
${\cal M}_{  \{ {\cal A}^I , {\cal S}_I \} }$ we have
fixed only the topology of the surfaces and their
intersection numbers with the loops.  There remain
a countable set of
diffeomorphism equivalence classes with
these specifications; they are
distinghished by the knotting of the loops and their
linking with each
other.

Of this remaining information, a certain
amount may be said to correspond to
information about the spacial geometry that
cannot be resolved by
measurements made using the quantum
coordinate system $\cal T$.  We
may imagine further refining the quantum
reference system by introducing
new surfaces by subdividing the tetrahedra in $\cal T$.
If we consider how
this may be done while keeping the toplogical relations
of the loops with
themselves and with the original set of surfaces fixed,
we may see that there
is a sense in which we can obtain a more precise
measurement of the
spatial quantum geometry associated with the topology
of the loops,
$\gamma$.  Of course, there is also a danger that by
subdividing too much
we may reach a point where additional surfaces
tell us nothing more about
the quantum geometry; the information about the
matter state and the quantum
geometry is entangled and cannot be easily
separated.   In a
further work, I hope to return to the problem
of how to disentangle the
geometrical from the matter information in such
measurements.

At the same time, it is clear that there is
information in the topology of the
loops that is not about the spatial geometry
and so cannot be resolved by
further refinement of the simplex based on the
matter state.  This includes
information about the routings through
intersections.  It is clear from this
and earlier considerations that the routings
through the intersections carry
information about the degrees of freedom
conjugate to the three geometry.

One can obtain this conjugate
information by measuring the
operators $T^I \equiv T[\partial {\cal S}_I$
defined
in the previous section.  If one measures
all $N$ of these operators, rather than the
$N$ areas, one determines the
parallel transport of the Ashtekar connection
around the edges of each
of the faces of the simplex $\cal T$.  Essentially,
this means
that one determines, instead of the areas of
the faces, the left handed curvatures evaluated
at the faces.  There is also a classical description
that can be associated with this measurement, it is described
in \cite{me-dieter}.

I would like to close this section by describing the sense in which
the results just described suggest an approach to a measurement
theory for quantum gravity\footnote{These remarks are enlarged
in \cite{me-dieter}.}.  The idea is to extend the principle
enunciated by Bohr that
what is observed in quantum mechanics must be
described in terms of
the whole system which includes a specification of both the atomic
system and the measuring apparatus.
In the case of quantum gravity,
the quantum system is no longer an atom, it is the whole spacetime
geometry.  As the quantum system is no longer
microscopic, but in fact
encompasses the whole universe, we can no
longer treat the measuring instrument classically while we treat
the spacetime geometry quantum mechanically.  Thus,
it is necessary
that
a measuring system that is to be used to determine
something about the spacetime
geometry must be prepared for the measurement
by putting it in some
definite quantum state.

In this paper I have described two conjugate sets of
measurements, which
determine either the areas of or the left handed
parallel transport around  areas a set of $N$ surfaces.
However, the  basic
features of the measurement process and how we
describe it should
extend to more general measurements.

Any measurement theory must have two components:
preparation and
measurement.
If we are to use this measuring instrument to
probe the quantum
geometry, we must prepare the quantum state of the
measuring instrument appropriately.  As we are
interested in describing
the theory at a diffeomorphism invariant level, we must give a
diffeomorphism invariant specification of the quantum state of
the measuring instrument such that, when we act on the combined
gravity matter state with the area operators we measure a set of
areas which are meaningful.

Now, the requirement of diffeomorphism invariance
forbids us from
preparing the measuring system in some state and
then taking the
direct product of that "appratus state" with a state
of the system.
Instead, the preparation of the measuring system must be
described by
restricting the system to an appropriate diffeomorphism invariant
subspace of the combined apparatus-gravity system.  Thus,
what I have done above is to prepare the quantum state of the whole
system in a way appropriate to the specification of the
measuring instrument by restricting the topological relations among
the $N$ surfaces so that they are faces of a given simplex $\cal T$.
This is done by restricting the quantum state of the system, prior to
the measurement, to be of the form (60).  After we have made this
restriction, we can be sure that the results of the $N$ measurements
will be a set of $N$ areas that can be ascribed to the faces of the
simplex $\cal T$.  Thus, the result of the measurement of the area
operators on prepared states of the form (60) is to produce a partial
description of the spatial geometry which is given by the piecewise
flat manifold ${\cal M}_{\gamma,{\cal S}_I} $.

Now, the $N$ area operators $\hat{A}^I$ commute with each other,
but they do not make a complete set of commuting
observables.  This is
because to each such peicewise flat manifold, which encodes the
results of the $N$ observables, there are a
countably infinite number of
diffeomorphism invariant states in the subspace
${\cal S}_{\cal T}$
which are degenerate as far as the values of the
$\hat{A}^I$ are
concerned.

We would then like to ask whether we can add
operators to the $\hat{A}^I$
to make a complete set of commuting operators.
We certainly can extend
the set, by subdividing some or all of the tetrahedra in $\cal T$ to
produce a simplicial complex with more surfaces.  This would
correspond to introducing more matter fields, which would make
it possible to specify more surfaces whose area is to
be measured and
by so doing  make a more refined measurement of
the quantum geometry.
But, notice that there is a natural limit to how much
one can refine one's
observations of the quantum geometry because one
can never measure
the area of any surface to be less than one-half  Planck area.

Further, note that no matter how large $N$ is, and no matter how
the $N$ surfaces are arranged topologically, there are always a
countably infinite set of states associated to each measurement of
the $\hat{A}^I$'s.  In this sense, it seems that one can
never construct
a physical measuring system that
suffices to extract all the information out of the quantum
gravitational field.  This is, of course, just a reflection of the fact that
the quantum gravitational field has an infinite number of degrees of
freedom, while any physical measuring instrument can
only record a finite amount of information.
However, note that we have come to an expression of this
fact in a way that is completely diffeomorphism invariant.  In
particular,
we have a characterization of a field with an infinite number of
degrees
of freedom in which we do not say how many degrees of freedom are
associated to each "point."  This is very good, as we know that no
diffeomorphism invariant meaning can be given to a point of space
or
spacetime.

Of course, there remains one difficulty with carrying
out this type of interpretation, which is that
the problem of time in quantum gravity
must be resolved so that we know how to speak of the time of the
observation.  In \cite{me-dieter} I show that
the resolution of the problem of
time may be carried out using
the ideas of spacetime diffeomorphism invariant
observables of
Rovelli\footnote{Note that such observables
 have been described in the $2+1$ case by
Carlip\cite{Carlip-time}, in the Gowdy
model by Husain\cite{viqar-time} and in the Bianchi-I model by
Tate\cite{tate-time}.}
\cite{problemoftime}.   To do this one constructs
the physical time dependent operators that correspond
to the
$\hat{A}^I$ and $T^I$.  These
depend on a time parameter $\tau$ which is the
reading of a physical clock built into the measuring instrument.
The $N$ operators $\hat{A}^I(\tau )$ will then commute with the
Hamiltonian constraint, and so act on physical states and their
eigenvalues return the values of the areas of the $N$ surfaces when
the physical clock reads $\tau$.    Given this, there seems to be no obstacle
to the observer
employing the projection postulate and
saying that the quantum state
of the matter plus gravity system is projected into a subspace of the
physical Hilbert space spanned by the appropriate eigenstates of
the $\hat{A}^I (\tau )$ and  that this is something that occurs just
after the measurement is made, in spite of the fact that she and her
apparatus are living inside the quantum system under study.

Thus,  despite various assertations to the contrary, there seems
to be no difficulty in applying a Copenhagen-like description of the
measuring process to the case of quantum cosmology in spite of the
fact
that the measuring instrument is inside the universe.  As long as we
can
prepare the measuring instrument in such a way that the quantum
state
of the whole matter-gravity state space is inside a subspace of the
state
space associated with a particular specification of the measuring
instrument
one can assign meaning to a set of commuting observations.
The implications of this are discussed further in
\cite{me-dieter}.

Finally, we may note that  one finds that the
$\hat{A}^I (\tau =0)$ are equal to the area operators constructed
in this paper\cite{me-dieter}, so that
the quantization of areas becomes a
physical prediction based on the spectra of a set of physical
operators of the theory.

\section{Conclusions}

I would like to close by making a number of comments about the
implications of the results obtained here.

1)  We see that in each case when we have succeded in
constructing the
definition of an operator in such a way that
when it is diffeomorphism invariant
it is automatically finite.  This is in accord with
the general arguments that
all spatially invariant diffeomorphism invariant
operators must be finite
that were given previously in \cite{review-lee,carloun}.
This suggests strongly that the
problem of constructing a finite theory of quantum gravity can be to
a great extent resolved at the diffeomorphism invariant
level.  The reason
is that once one imposes spatial diffeomorphism invariance
there is no
longer any physical meaning that can be given to a point in space.
As a result, although the theory still has an infinite
number of degrees
of freedom, in the sense discussed in section 6, it is no
longer meaningful
to speak of the field as having a certain number of
degrees of freedom
per point.  Instead, there seeems to be a natural limitation to how
many degrees of freedom there can be inside of a Planck
volume due to the
discreteness of the spectra of the geometrical operators that measure
area and
volume\cite{review-lee}.
This in
turn suggests that the problem
of finiteness has little to do with the dynamics of the
theory or the choice of matter couplings, which are
coded in the Hamiltonian
constraint.

2)  We also see that the conclusions of
previous analyses of the measurement
problem in quantum gravity by DeWitt and
others are confirmed.  The key
conclusion of these works was that it should
be impossible to meaningfully
resolve distance scales shorter than the Planck
scale.  We see that this is
the case here, because the possible values of physical
areas that can be
gotten from a diffeomorphism invariant measurement
procedure are
quantized in units of the Planck area.

We also see that any particular configuration of the
matter fields that are
used to define the reference system can only be used
to resolve a certain
finite amount of information about the space time
geometry.  This is a
consequence of using the quantum theory to describe
the reference system
as well as the gravitational field.  This is certainly consistent with
the general observation that diffeomorphism
invariant measurements are
about relations between the gravitational field and the measuring
instruments.  If we want a measurement system
which is able to resolve
$N$ different spatial distances, it had better come equiped with $N$
distinguishable components.

3)  We see that a large class of doubts about
the physical applicability of the
description of quantum states of geometry by means of the loop
representation can now be put to rest.  Note
that any spacial geometry in which the components of
the curvatures are small in Planck units can
be approximated by
a Regge
manifold in which the areas of the faces are
integral multiples of half the
Planck area.  As a result we see from the
correspondence between Regge manifolds and
quantum states arrived
at in section 6 that any such spatial geometry can be associated
with a diffeomorphism invariant
quantum state in the loop representation.  This allows
us to extend the
discussion of the classical limit of quantum gravity developed in
\cite{weave} to the diffeomorphism invariant level.

4)  It would be very interesting to be able to
characterize the quantum
geometry associated with  diffeomorphism invariant
states of the pure
gravitational field.  The results obtained with matter
fields as reference
systems suggest that there should be a basis of states
which are diagonal
in some set of diffeomorphism invariant operators which
measure the three
geometry and that this basis contains the characteristic
states of non-intersecting knots and links.  The problem is
to construct an
appropriate set of diffeomorphism invariant classical
observables which
are functions only of the gravitational field and translate
them into quantum
operators while preserving the diffeomorphism invariance.  We
already know how to construct a few such operators,
which measure
the areas of
extremal surfaces and, in the case that it is spatially compact,
the volume
of the universe\cite{review-lee}.

One approach to the construction of such observables
could be by mimicking
the results of this paper by constructing observables that measure
the areas
of surfaces on the faces of a given simplex, and asking that all
the areas are
extremized as the whole simplex is moved around in the geometry.
Constructions along this line are presently under study.

5)  Given the present results, a new approach to the construction
of the full
dynamical theory becomes possible.  This is to impose an inner
product consistent with the reality conditions at the diffeormorphism
invariant level and then project the Hamiltonian constraint into the
resulting Hilbert space of diffeomorphism invariant states.
The physical
state space would then be found as a subspace of the space of
diffeomorphism invariant states.  constraint The
main difficulties facing such an approach are the problem
of expressing
both the reality conditions and the Hamiltonian constraint in
diffeomorphism
invariant forms.

6)  As I commented above, the form of the Hamiltonian constraint (7)
for gravity coupled
to the simple, massless antisymmetric tensor gauge
field is particularly simple.
It would be very interesting to see if solutions to the
Hamiltonian constraint
for the coupled gravity matter system could be obtained,
if not exactly, in the
context of some perturbative expansion.  It is intersting to
note that exact solutions
can be obtained in the strong coupling limit in which $k$ is
taken to zero
(because $k$ is inverse to what would usually be written
as the coupling
constant).  In
this limit, only the second term of (7) survives.  It is easy
to show, using
a regularization of the type introduced in \cite{carlolee}
that one then has a class of solutions
of the form $\Psi [\{ {\cal S},\gamma \}]$ in which the loops
never intersect the
surfaces or in which the loops always lie in the surfaces.
It would be very
interesting to then develop a strong coupling expansion
to construct
approximate expressions for solutions
for finite $k$.   It would also be very interesting to see if
one could recover
from the semiclassical limit of the gravity matter system
the solutions to the
Schroedinger equation described in \cite{rodolfo-antisym}.

7)  It is interesting to note that surfaces play an
interesting role in two
mathematical developments connected to the Ashtekar variables
and the loop representation.  In \cite{baez}
Baez extends the loop representation to the case in which the
spatial manifold has a boundary, and shows that in this
case there is an interesting
algebra of operators that acts on the diffeomorphism
invariant states.  In
\cite{catalouis}, Crane proposes a new interpretative
scheme for quantum gravity
in which Hilbert spaces of states coming from conformal
field theories are defined
on surfaces which are identified with observers
and measuring instruments.  Both
proposals need to be completed by the construction
of explicit diffeomorphism
invariant observables associated to surfaces, and it
would be interesting
to see if  the operators
described here can thus play a role in these proposals.

8)  Finally, I would like to address the issue of the
use of $N$ separate
matter fields to label the operators that measure the
areas of the $N$
surfaces.   This is clearly necessitated by the idealization
in which I use
the values of a field to specify a set of physical
surfaces in a very simple
way.  The point is that there must be a physical
way to distinguish the
$N$ different surfaces in terms of the configurations
of the matter fields.
In real life, in which measuring instruments of
arbitrary complexity are
constructed from a small number of fields there is
no difficulty with specifying quantum states associated
to some degree of
precision with an arbitrary number and configuration
of surfaces.  In
a realistic situation the configuration is complex
enough to allow an
intrinsic labeling of the different
surfaces\footnote{This accords with the observation
stressed by Barbour that real physical observables are
well defined because the world is sufficiently complex to
allow the events of spacetime to be distinguished
by the values of the physical fields \cite{julian-heap}.}  Of course,
another issue also
arises when we construct the surfaces out of
realistic physical fields,
which is
that there will be restrictions to the accuracy of the measurements of
thea
areas due to the fact that matter is made out of atoms.  It is not,
however,
impossible that such limitations can be overcome by a clever use of
matter and other fields to specify very small surfaces.  What the
present
results suggest, however, is that no
matter how clever we are with the design
of our measuring instruments, it will be impossible to
measure the area of
any physical surface to be less than half the Planck area.

\section*{ACKNOWLEDGEMENTS}

I would like to thank Abhay Ashtekar, Rodolfo Gambini and
Carlo Rovelli for
critical readings
of a draft of this manuscript.  I am very grateful to  them and to
Julian Barbour and Louis Crane for crucial
conversations about this work and
the general problem of constructing
diffeomorphism invariant observables.  I would also like
to thank Alan Daughton and Rafael Sorkin for conversations about
simplicial manifolds.
This work was supported by the National
Science Foundation under grants PHY90-16733 and
INT88-15209 and
by research funds provided by Syracuse University.


\begin{thebibliography}{99}
\bibitem{alan-personal}I would like to thank Alan Daughton for a
helpful
conversation about this.
\end{thebibliography}
\end{document}